\documentclass[twocolumn,showpacs,preprintnumbers,pre]{revtex4}
\usepackage{graphicx,epsfig}
\usepackage{dcolumn}
\usepackage{bm}
\begin{document}
\title{Motion of the hydrogen bond proton in cytosine\\
and the transition between its normal and imino states}
\author{Zhen-Min Zhao$^2$, Qi-Ren  Zhang$^{1,2}$,  Chun-Yuan Gao$^2$ and
  Yi-Zhong Zhuo$^3$\\
$^1$CCAST(World Lab),P.O.Box 8730,Beijing,100080, China\\
$^2$Department of Technical Physics, Peking University, Beijing
100871, China
\\ $^3$China Institute of Atomic
Energy, Beijing 102413, China}

\begin{abstract} The potential energy surface of the H13 proton
in base cytosine
of the DNA molecules is calculated {\it ab initio} at the Gaussian98
MP2/6-311G(d,p) level. Two potential wells are found. One
corresponds to the normal cytosine, while the other corresponds to
its imino tautomer. The bindings of the proton in these wells are
stable enough against the thermo-disturbance. The motions of the
proton  in these wells are oscillations around the nearest nitrogen
atom like the pendula, and may move far away from the nitrogen atom
to form the hydrogen bond with other bases. The estimated tunneling
probability of the H13 proton from one well to another well shows
that the life time of the proton staying in one of these wells is
about  6$\times10^2$ yr. It is too long to let tautomers of cytosine
be in thermodynamical equilibrium in a room temperature gas phase
experiment. The biological significance of these result is
discussed.

\bigskip\noindent {\bf Keywords}:Hydrogen bond proton, Potential
energy surface, Tunneling
\end{abstract}
\pacs{87.14.Gg, 87.15.-v}
\maketitle
\section{Introduction}
 More than 40 years ago, Per-Olov Lowdin \cite{1}
proposed a program for the research of mutation, aging, and cancer
formation on the basis of  DNA changes due to the tunneling of the
hydrogen bond proton through the barrier separating its normal and
ill configurations. It was unable to realize this attractive program
at that time, since both the theoretical chemical method and the
computational ability were not enough developed. Having {\it ab
intio} method with its fairly developed software package and high
performance computers in hand, one is able to try it now. Many works
have been done along this direction. References \cite{2}-\cite{15}
are some earlier examples. However, we do not yet see the previous
report on the {\it ab initio} calculation about the potential
surface and the corresponding motion of the hydrogen bond proton.

The four bases (A, T, C, G) in a DNA molecule usually stay in their
normal configurations, but occasionally transfer to their unusual
imino/enol configurations with small probabilities. It is the
transitions A$\leftrightarrow \mbox{A}_{\mbox{imino}}$,
T$\leftrightarrow \mbox{T}_{\mbox{enol}}$, G$\leftrightarrow
\mbox{G}_{\mbox{enol}}$, and C$\leftrightarrow
\mbox{C}_{\mbox{imino}}$. The imino/enol tautomers will be paired in
the way $\mbox{A}_{\mbox{imino}}\cdots$C,
A$\cdots\mbox{C}_{\mbox{imino}}$, $\mbox{G}_{\mbox{enol}}\cdots$T,
and G$\cdots\mbox{T}_{\mbox{enol}}$. It makes the possible changes
\begin{equation}\mbox{C}\cdots\mbox{G}\leftrightarrow
\mbox{T}\cdots\mbox{A}\end{equation}of the base pairs in DNA during
its double replications. For an example, the transition
C$\rightarrow$ C$_{\mbox{imino}}$ changes the pair C$\cdots$G  to
C$_{\mbox{imino}}$ G , and the pairing
A$\cdots\mbox{C}_{\mbox{imino}}$ produces  pairs
A$\cdots\mbox{C}_{\mbox{imino}}$ and  G$\cdots$C in a first
replication. A further replication produces pairs T$\cdots$A,
C$_{\mbox{imino}}\cdots$A ,C$\cdots$G , and C$\cdots$G. In the first
case, a pair T$\cdots$A appears at the position originally occupied
by the pair C$\cdots$G . This is a change of heredity code,
therefore is serious. It may influence the development and
reproduction of a living system, cause aging, cancer, and mutation
for examples. To understand and research these kinds of processes at
the level of quantitative theoretical physics would be highly
desirable.
\begin{figure}[h]
\centerline{\epsfig{file=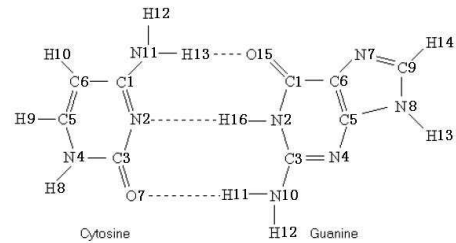,width=8cm}}

\caption{The bases cytosine and guanine and their pairing by
hydrogen bonds (dashed lines) The attached number is for specifying
the atom in a base}
\end{figure}

\begin{figure}[h]
\centerline{\epsfig{file=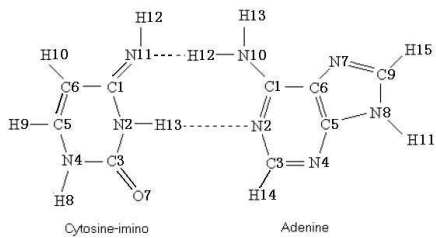,width=8cm}}

\caption{The bases cytosine-imino and adenine and their pairing by
hydrogen bonds} \end{figure}

Figure 1 and 2 show the structure formulae of base cytosine and its
imino tautomer, as well as their different pairings. The difference
between these two tautomers of cytosine lies mainly in the different
positions of the proton H13. This difference in turn causes the
different distribution of hydrogen bonds, and finally causes the
different pairings. Therefore, to clarify the potential surface of
proton H13 and its motion under the influence of this potential is
the key for understanding the transition C$\rightarrow$
C$_{\mbox{imino}}$ between cytosine tautomers and the change
C$\cdots\mbox{G}\rightarrow \mbox{T}\cdots$A of the base pair in
DNA.

Section 2 is devoted to finding the potential surface of the proton
H13 (the hydrogen bond proton) for base cytosine by {\it ab initio}
calculation. In section 3, we prove that the proton H13 is bound in
both, normal and imino, configurations, and it is ready for forming
the right hydrogen bond of pairing. In section 4 we calculate the
tunneling probability of the proton H13 between these two
configurations and discuss its biological significance. Section 5
contains the conclusion and a discussion of the reliability of the
results.
\section{Potential energy surface of the proton H13 for
cytosine} We have calculated the single point energy (SPE) of the
proton H13 {\it ab initio} by MP2 method in Gaussian98 package with
the 6-311G(d,p) base set. For every given position of this proton,
positions of other atoms in cytosine are optimized by the
minimization of the energy of the electron system. It causes a small
change of these positions in cytosine. Using a  coordinate system
fixed on the known structure formula of cytosine, the obtained map
of the potential energy surface of the proton H13 is shown in figure
3. In this figure,  the origin of the  coordinate system is at the
position of N11 in figure~1, the $x$ axis is along its direction
N11-C1, and the $y$ axis is perpendicular to $x$ axis and points to
the right. The bluer is the point in figure 3 the lower is the
potential at that position. The number attached on the line shows
the single point energy along the contour. We see clearly two
potential wells. The left one corresponds to the cytosine, and the
right one corresponds to its imino tautomer. It explains the
structure of these two cytosine tautomers shown in figures 1 and 2.
\begin{figure}[h]
\centerline{\epsfig{file=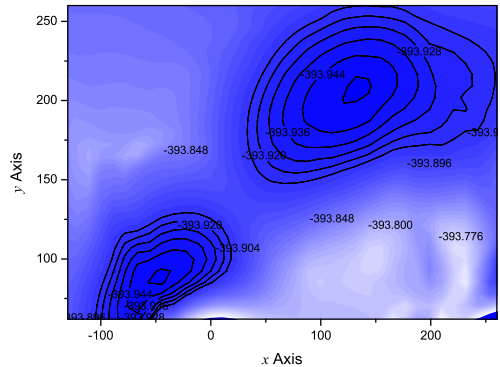,width=8cm}}

\caption{Contour diagram for the potential energy of the proton
H(13) in cytosine. The energy is in atom unit (a.u.), and the length
is in the unit of pico-meter (pm). }\end{figure}

\section{Binding of proton H13 in the base cytosine and its imino
tautomer} Around the bottom of a well, we fit the calculated single
point energies by a quadratic form of coordinates and find the
minimum of SPE by differentiation. The minimum SPE of the left well
is $-$393.9535268 a.u and that of the right well is $-$393.952654
a.u. Moving the origin of the coordinate system to the position of
minimum SPE, the terms containing first degree of coordinate
disappear. Finding the principal axes of the coordinate system by
the standard method, we may transform the quadratic form into a sum
of squares of normal coordinate. For the left well, we take a
coordinate system ($\xi$, $\eta$) with the $\xi$ axis along the
direction N11-H13 in cytosine of figure 1, and $\eta$ axis
perpendicular to it. The normal coordinates ($X$, $Y$) is related to
($\xi$, $\eta$) by a rotation.
\begin{equation}
\left(\begin{array}{c}X\\
Y\end{array}\right)=\left(\begin{array}{cr}0.94&\;\;-0.34\\
0.34&\;\;0.94\end{array}\right)\left(\begin{array}{c}\xi\\
\eta\end{array}\right)\; .\label{4}\end{equation} The SPE as a
function of normal coordinates ($X$, $Y$) is
\begin{equation}E=-393.9535+0.088937Y^2\; ,\end{equation}in which the single point
energy $E$ is in unit of a.u., and the unit of the length is
angstrom. We see that $E$ is independent of the coordinate $X$ at
the bottom of this well. The potential surface here looks like a
valley along $X$ direction. The transformation (\ref{4}) shows it is
a direction near the line N11-H13. Along this direction, the proton
is easier to go far away from the atom N11 to form a hydrogen bond
with O15 in guanine. It explains the pairing shown in figure~1. The
oscillation is along the $Y$ direction perpendicular to this
direction. It is an oscillation around the atom N11 in cytosine,
like a pendulum `hang' at N11. The frequency is
\begin{eqnarray}\nu&=&\left.\sqrt{\frac{2\times0.088937\times27.21\times1.60217733
\times10^{-19}}{1.67231\times10^{-27}\times\left(10^{-10}\right)^2}}\right/(2\pi)
\nonumber\\&=&3.427\times10^{13}(\mbox{Hz})\; ,\end{eqnarray} and
the zero point energy (ZPE) is
\begin{equation}\frac12h\nu=0.0026\mbox{a.u}\; .\end{equation}
Adding this ZPE with the SPE at the bottom of the well, we find that
the energy of the proton H13 is $-$393.9509 a.u.. It is lower than
the lowest maximum SPE $-$393.8895 a.u. around the well. The proton
H13 is therefore bound in the normal cytosine, with a binding energy
$b$=0.0614a.u..

For the right well in figure 3, we take the coordinate system ($x'$,
$y'$), with the $x'$ axis parallel the $x$ axis in figure 3, while
its origin is moved to the bottom of the well. The normal
coordinates ($X$, $Y$) is related to ($x'$, $y'$) by a rotation

\begin{equation}
\left(\begin{array}{c}X\\
Y\end{array}\right)=\left(\begin{array}{cr}0.30&\;\;-0.95\\
0.95&\;\;0.30\end{array}\right)\left(\begin{array}{c}x'\\
y'\end{array}\right)\; .\label{8}\end{equation} The SPE as a
function of $X$ and $Y$  is
\begin{equation}E=-393.9527+0.12878Y^2\; .\end{equation}
$E$ is independent of the coordinate $X$  again at the bottom of the
well, and the potential surface looks like a valley along the $X$
direction once more. The transformation (\ref{8}) shows it is a
direction almost perpendicular to the line N11-C1. Along this
direction, the proton is easier to go far away from the atom N2 of
cytosine-imino in figure 2 to form a hydrogen bond with atom N2 in
adenine. It explains the pairing shown in figure 2.  The oscillation
is almost along this line. It is an oscillation around the atom N2
in cytosine-imino, like a pendulum `hang' at this atom N2. The
frequency is
\begin{eqnarray}\nu&=&\left.\sqrt{\frac{2\times0.12878\times27.21
\times1.60217733\times10^{-19}}{1.67231\times10^{-27}\times
\left(10^{-10}\right)^2}}\right/(2\pi)\nonumber\\&=&4.12\times10^{13}(\mbox{Hz})\;
,\end{eqnarray} and the ZPE is
\begin{equation}\frac12h\nu=0.00313\mbox{a.u}\; .\end{equation}
The lowest maximum SPE around two wells are at the same point, it is
a point between these two wells. The SPE $-$393.8895 a.u. at this
point is called the energy of the transitional state. The sum of the
ZPE and the bottom SPE is the energy $-$393.9496 a.u. of the proton
H13 in this well. It is also lower than the transitional state
energy. The proton H13 is bound too in the cytosine-imino. The
binding energy here is $b'$=0.0601 a.u.

The thermo-energy  in a living system may be estimated to be
$k_BT$=0.001a.u. at the room temperature $T\simeq300$K. We see that
$b$ and $b'$ are much larger than the thermal energy. It implies
that both cytosine and its imino tautomer are stable against the
thermal disturbance.

\section{Tunneling between two tautomers and its biological
significance}
   According to quantum mechanics, there is still a possibility of
   the proton tunneling from one well to another well.
   We first find the path of lowest potential connecting the centers of
   these two wells. It is described by the equation
\begin{eqnarray}y(x)&=&1.1417352+0.71244583x+0.3508047x^2\nonumber\\&+&0.04422011x^3
+0.06419875x^4-0.56534255x^5\nonumber\\&+&0.25248558x^6\end{eqnarray}in
the coordinate system used in figure 3, with the length unit being
changed to angstrom. The potential energy in atom unit along the
path is fitted by the formula
\begin{eqnarray}V(x)&=&-393.908877+0.103691x-0.102525x^2\nonumber\\&-&0.3063312x^3
+0.102250x^4+0.355250x^5\nonumber\\&-&0.195806x^6\; .\end{eqnarray}
The potential barrier between two wells of the proton H13 is shown
in Figure 4. According to the WKB formula, the penetration
probability of the proton through this barrier is
\begin{eqnarray}W&=&\exp\left\{-\frac2\hbar\int_s\sqrt{2m_p\left[V(x)-E\right]}\,
{\rm d}l\right\}\nonumber\\&=&\exp\left\{-\frac2\hbar\int_a^b
\sqrt{2m_p\left[V(x)-E\right]\left[1+\left(\frac{{\rm d}y}{{\rm
d}x}\right)^2\right]}\;{\rm
d}x\right\}\nonumber\\&\approx&\exp[-54.85]
\approx1.50929\times10^{-24}\; .\end{eqnarray} Here $E=-$393.9496
a.u. is the total energy of the proton H13 in cytosine-imino, $a$
and $b$ are the $x$ coordinates of the proton to be determined by
the condition $E=V(x)$  with $b>a$, and $s$ is the lowest potential
path terminated by points $x=a$ and $x=b$.
\begin{figure}[h]
\centerline{\epsfig{file=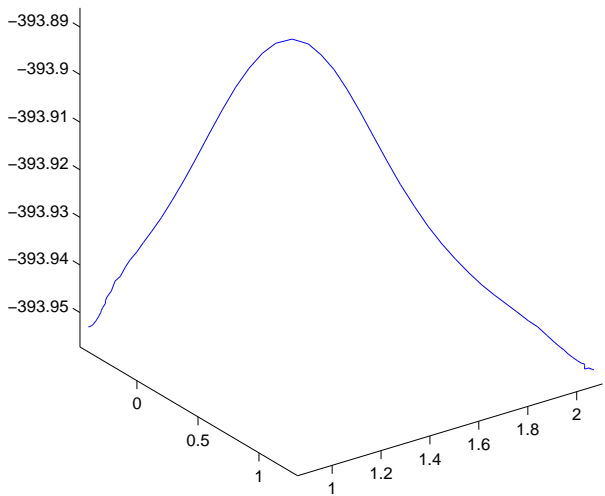,width=8cm}}

\caption{The potential barrier between two wells of the H13
proton}\end{figure}

The penetration probability in unit time is
\begin{eqnarray}P&=&\nu W=3.427\times10^{13}\times 1.50929\times10^{-24}
\mbox{Hz}\nonumber\\&=&5.168\times10^{-11}\mbox{s}^{-1}\label{15}\end{eqnarray}
Therefore, the mean life time for a proton staying in one of these
wells is 6$\times10^2$ years. It means the time needed for reaching
the thermo-equilibrium is more than six hundred years. One could not
see this equilibrium of gas type cytosine and its imino tautomer in
his one hundred year life. We had better to assume that the large
ratio between populations of cytosine and its imino tautomer in
nature is not a result of thermo-equilibrium, but rather the natural
selection resulting in biological evolution. From (\ref{15}) we also
see, in one's one hundred year life time, about sixteen percent of
his cytosine bases may have spontaneously transferred into some
imino form. Some of them may transfer back into normal form of
cytosine, while some others, according to Lowdin's conjecture, may
induce aging, mutation or cancer in further replications of DNA. Of
course, cytosine as well as other bases of DNA molecules are not
isolated in living systems. Their transitions to `ill' tautomers
must be influenced by their environment. To clarify this influence,
one has to do more complicated calculations for a base-environment
system. A deeper research along this direction would be interesting.
\section{Conclusion and discussion}
The {\it ab initio} quantum mechanical calculation reproduces the
structures of cytosine and its imino tautomer, including the
possible positions and directions of their hydrogen bond formation.
It makes us be confident that this method is applicable and
effective in the bio-physics research, and predict the transition
probability between normal and imino states of cytosine. The result
is reasonable. Its biological significance is discussed above and
worthy to explore further.

There are a large number of atoms in a bio-molecule. The energy of
the molecule is contributed by atoms in it and by their
interactions. The energy change of the molecule in the physical,
chemical or biological reactions is usually a tiny part of the total
energy of the molecule. To study the behavior and the energy change
of the bio-molecule in these reactions, we need very precise
calculations to keep many digits of the value be correct. For an
example, as we have seen at the beginning of section 3, to compare
the single point energy of the proton at the centers of two wells,
we need at least first 6 digits of the calculated values be precise.
This is not easy in such a huge amount of numerical calculations.
For the results reported here, the general tendency of the potential
surface may be reliable, but the difference of SPE for specially
chosen pair of points may not be so, since the first 5 most reliable
digits are subtracted. On the other hand, the tunneling probability
may be reliable, since a part of statistical errors may be canceled
by each other in an integration. Anyway, to develop the more precise
programs with specified upper bound of possible errors would be very
desirable in further researches along this direction.

This Work was supported by National Nature Science Foundation
Committee of China with grant numbers 10305001 and 10475008, and by
the High Performance Computing Center of China (Beijing).

\end{document}